\documentclass[a4paper]{article}

\usepackage{INTERSPEECH2020}
\usepackage{multirow}
\usepackage{makecell}
\usepackage[flushleft]{threeparttable}
\usepackage{hyperref}

\title{Cross-Language Transfer Learning, Continuous Learning, and Domain Adaptation for End-to-End Automatic Speech Recognition}
\name{Jocelyn Huang$^1$, Oleksii Kuchaiev$^1$, Patrick O'Neill$^2$, Vitaly Lavrukhin$^1$,\\
Jason Li$^1$, Adriana Flores$^1$, Georg Kucsko$^2$, Boris Ginsburg$^1$
\thanks{Preprint. Submitted to INTERSPEECH.}
}

\address{
$^1$NVIDIA, Santa Clara, CA, USA\\
$^2$Kensho, Cambridge, MA, USA}
\email{\{jocelynh,okuchaiev,vlavrukhin,jasoli,adrianaf,bginsburg\}@nvidia.com,
\{patrick.oneill,georg\}@kensho.com
}

\begin{document}

\maketitle

\begin{abstract}
In this paper, we demonstrate the efficacy of transfer learning and continuous learning for various automatic speech recognition (ASR) tasks. We start with a pre-trained English ASR model and show that transfer learning can be effectively and easily performed on: (1) different English accents, (2) different languages (German, Spanish and Russian) and (3) application-specific domains. Our experiments demonstrate that in all three cases, transfer learning from a good base model has higher accuracy than a model trained from scratch. It is preferred to fine-tune large models than small pre-trained models, even if the dataset for fine-tuning is small.
Moreover, transfer learning significantly speeds up convergence for both very small and very large target datasets.

\end{abstract}


\section{Introduction}

End-to-end training of Automatic Speech Recognition (ASR) models requires large datasets and heavy compute resources. There are more than 5,000 languages around the world, but very few languages have datasets large enough to train high quality ASR models \cite{Wang2015}. Additionally, datasets for different domains are also highly imbalanced with respect to domain-specific vocabulary, even for English.
Transfer Learning (TL) is one of the most popular methods used to build ASR models for low resource languages from a model trained for another language \cite{Schultz2001}, based on the assumption that phoneme representation can be shared across different languages.
TL can be also used for adaptation of a generic ASR model to a more narrow domain, such as medical or financial reports, for the same language. Continual learning is a sub-problem within TL wherein models that are trained with new domain data should still retain good performance on the original source domain.

In this paper we demonstrate good results for transfer learning across accents, domains, and even languages. We apply the same simple recipe in all experiments. First, we take an encoder from the QuartzNet model \cite{Quartznet} pre-trained on 3,300 hours of public English data. If a target domain has a different alphabet, we randomly initialize a new shallow decoder. Otherwise, we load a pre-trained decoder as well. Then we fine-tune the resulting network on a target dataset starting with an $\approx 10$x smaller learning rate than what was used to pre-train the model from scratch.
We found this recipe works whether the fine-tuning dataset is smaller than the dataset used for pre-training, the same size, or even substantially larger.  Interestingly, final performance is always better when starting from the pre-trained model vs training on all of the data from scratch. This holds even in the case where the fine-tuning dataset is an order of magnitude larger than the pre-training dataset.  We also show that catastrophic forgetting will happen unless some of the pre-training data is also included in the fine-tuning step.  These results are consistent across all our experiments.\footnote{
All experiments were performed using the NeMo toolkit \cite{nemo2019}.}

\section{Related work}

\subsection{Cross-language transfer learning}

Transfer learning for ASR was originally used for Gaussian Mixture Model - Hidden Markov Model (GMM-HMM) systems, based on the idea that phoneme representation can be shared across different languages. Anderson et al \cite{Anderson1994} applied this idea to acoustic modeling using the International Phonetic Alphabet (IPA). The cross-language acoustic model adaptation was explored in depth in the GlobalPhone project \cite{Schultz2001}. It was based on two methods: (1) partial model adaptation for languages with limited data, and (2) boot-strapping, where the model for new target is initialized with a model for other language and then completely re-trained on target dataset.
TL was also used for hybrid Deep Neural Network (DNN) - HMM models \cite{Wang2015}. The basic idea was that the features learned by DNN models tend to be language-independent at low layers, so the low-level layers can be shared by all languages \cite{Huang2013}. This hypothesis was experimentally confirmed by TL between ASR models for Germanic, Romance, and Slavic languages \cite{Swietojanski2012,Ghoshal2013}.

Kunze et al \cite{Kunze2017} applied TL to DNN-based end-to-end ASR models, and adapted an English ASR model for German. In their experiments they used a Wav2Letter model and froze the lower convolutional layers while retraining the upper layers.
Similarly, Bukhar et al \cite{Bukhari2017} adapted a multi-language ASR model for two new low-resource languages (Uyghur and Vietnamese) by retraining the network's last layer.
Tong et al \cite{Tong2017} trained \textit{multilingual CTC-based model} with IPA-based phone set, and then adapted it for a language with limited data. They compared three approaches for cross-lingual adaptation: (1) retraining only an output layer; (2) retraining all parameters; (3) randomly initializing weights of the last layer and then updating the whole network. They found that updating all the parameters performs better than only retraining the output layer.

\subsection{Transfer learning for domain adaptation}

Ueno et al \cite{Ueno2018} used TL to adapt a generic ASR model to new domains: call center and voice search tasks. They used a model similar to Listen-Attend-Spell \cite{chan2015}. They compared three methods: (1) training with the target data from scratch, (2) freezing encoder and retraining the decoder, and (3) updating all parameters.
Moriya et al \cite{Moriya2018} suggested progressive knowledge transfer to adapt ASR model to two new environments (in-car and distant talk) while keeping the pre-trained knowledge. The basic idea is to add an additional `column' to the model for each new domain. The new column is trained with only the new domain the while other columns are frozen.

\section{Model architecture}

\begin{figure}[htb!]
 \centering
 \includegraphics[width=0.8\linewidth]{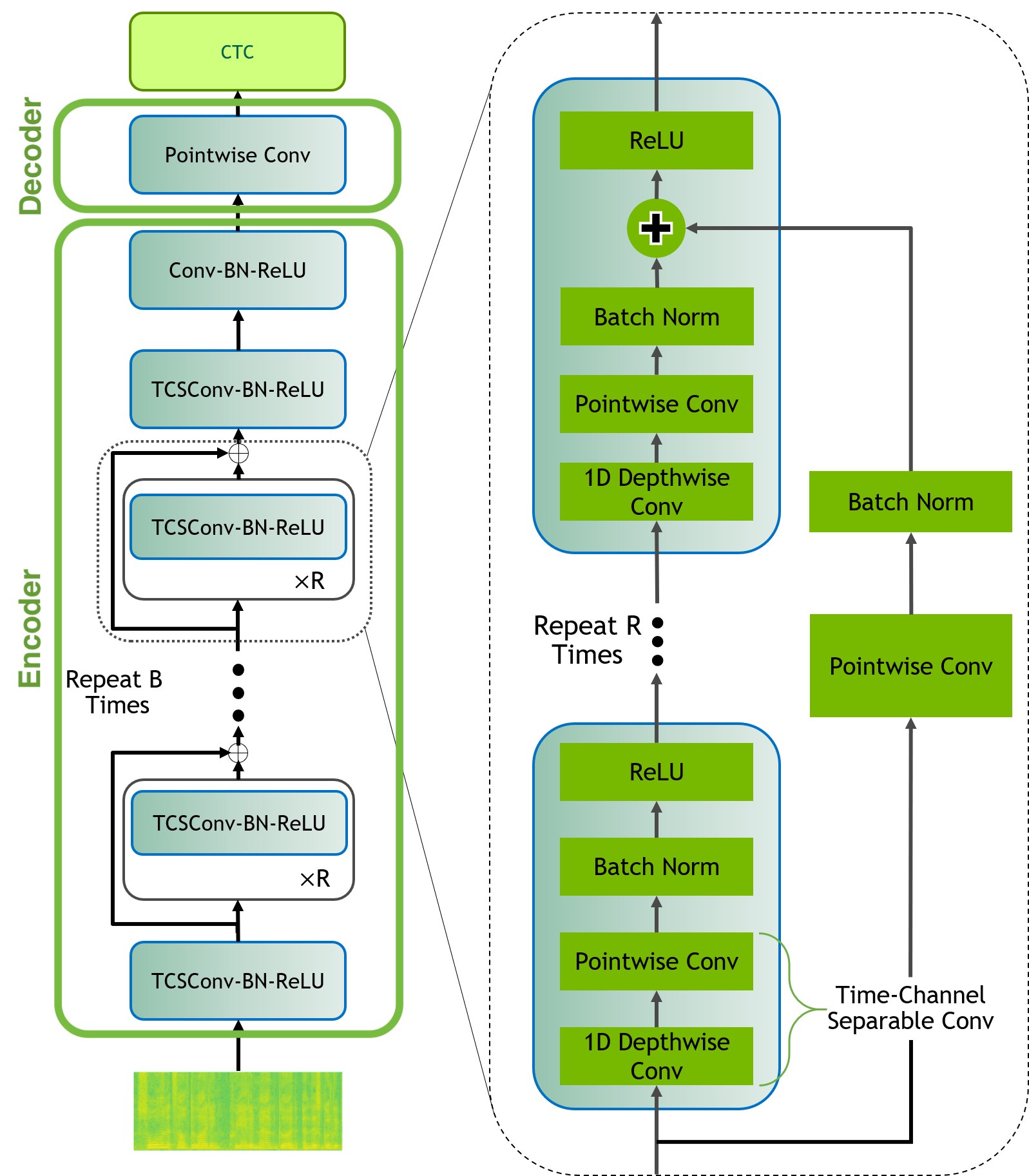}
 \caption{QuartzNet BxR architecture \cite{Quartznet}}
 \label{fig:quartz_arch}
\end{figure}

In our experiments, we use a QuartzNet  model trained with Connectionist Temporal Classification (CTC) loss \cite{graves2006}. QuartzNet employs 1D time-channel separable convolutions, a 1D version of depthwise separable convolutions \cite{Chollet2017}. Each 1D time-channel separable convolution block can be separated into a 1D convolutional layer with kernel length $K$ that operates on each channel separately \textit{across $\pmb{K}$ time frames} and a point-wise convolutional layer that operates on each time frame independently \textit{across all channels}.

QuartzNet models have the following structure: they start with a 1D time-channel separable convolutional layer $C_1$ followed by a sequence of blocks (see Figure.\ref{fig:quartz_arch}).
Each block $B_i$ is repeated $S_i$ times and has residual connections between each repetition. Each block $B_i$ consists of the same base modules repeated $R_i$ times. The base module contains four layers: 1) $K$-sized depthwise convolutional layer with $C$ channels, 2) a pointwise convolution, 3) a normalization layer, and 4) ReLU. The last part of the model consists of one additional time-channel separable convolution ($C_2$), and two 1D convolutional layers ($C_3, C_4$). The $C_1$ layer has a stride of 2, and $C_2$ layer has a dilation of 2.
Table \ref{tab:Model_Architectures} describes the QuartzNet 15x5 model. There are five unique blocks: $B_1$ - $B_5$, and each block is repeated $S=3$ times ($B_1-B_1-B_1-...-B_5-B_5-B_5$).

\begin{table}[htb!]
\caption{
QuartzNet 15x5. The model has 5 groups of blocks. Blocks in the group are identical, each block $B_k$ consists of \textbf{R} time-channel separable \textbf{K}-sized convolutional modules with \textbf{C} output channels. Each block is repeated \textbf{S} times.
}

\vspace{4pt}
\label{tab:Model_Architectures}
\centering
\scalebox{0.8}
{
\begin{tabular}{c c c c c c c}
 \hline
   \textbf{Block} & \textbf{R} & \textbf{K} & \textbf{C}& \textbf{S} \\
 \hline
 $C_1$ & 1 & 33 & 256 & 1\\
 \hline
 $B_1$ & 5 & 33 & 256 & 3\\
 $B_2$ & 5 & 39 & 256 & 3\\
 $B_3$ & 5 & 51 & 512 & 3\\
 $B_4$ & 5 & 63 & 512 & 3\\
 $B_5$ & 5 & 75 & 512 & 3\\
 \hline
 $C_2$ & 1 & 87 & 512 & 1\\
 $C_3$ & 1 & 1 & 1024 & 1\\
 $C_4$ & 1 & 1 & $\|$labels$\| $   & 1\\
 \hline
 \textbf{Params, M}& & & & 18.9\\
 \hline
\end{tabular}
}
\end{table}

\section{Transfer learning for English accents}

In this section we describe TL experiments with two English accented  datasets: the  Singapore National Speech Corpus (NSC) \cite{Koh2019} and the   Corpus of Regional African American Language (CORAAL) \cite{coraal}. For both datasets we observe consistent improvements in accuracy using TL versus training from scratch.

\subsection{Singapore English}

National Speech Corpus (NSC) is a large Singapore English speech dataset \cite{Koh2019}. In this work we used two parts of NSC V2.02: NSC1 - phonetically balanced read speech, and NSC2 - read speech with local named entities.
There is no official train/validation split for NSC. For NSC1, we randomly selected 50 speakers for validation, and removed all phrases that were spoken by validation speakers from the train data.
For NSC2, we followed a similar procedure with the only difference being that for validation we selected 31 speakers from NSC1 and 19 new speakers from NSC2. Thus, train and validation sets are disjoint both speaker-wise and phrase-wise (see Table.~\ref{tab:NSC}).

\begin{table}[th]
\centering
\caption{National Speech Corpus: train and validation split.}
\label{tab:NSC}
\scalebox{0.9} {
\begin{tabular}{l c c c}
 \toprule
 \textbf{Dataset} & \textbf{\# speakers} & \textbf{\# utterances} & \textbf{\# hours} \\
 \midrule
 NSC1 train & 987 & 1,188,390 & 1,856 \\
 NSC1 val & 50 & 108,246 & 159 \\
 NSC2 train & 981 & 2,084,076 & 2,691 \\
 NSC2 val & 50 & 121,386 & 160 \\
 \bottomrule
\end{tabular}
}
\end{table}

Since NSC is rather a large dataset, all experiments were done using QuartzNet 15x5 with fine-tuning for 10 epochs only.
We first trained three models on  non-accented datasets. The first model was trained on LibriSpeech (LS) only \cite{panayotov2015librispeech}. The second model was trained on LibriSpeech and Mozilla Common Voice (LS+CV) \cite{ardila2019common}. The third model (5D) was trained on five combined datasets: LS, CV, Wall Street Journal (WSJ)~\cite{wsj}, Fisher~\cite{fisher}, and Switchboard~\cite{Switchboard}. We evaluated all three models without fine-tuning on our NSC1 validation set. As expected, with more training data we observed lower word error rate (WER) on both LS and NSC1 (first three rows of Table.~\ref{tab:NSC1}).

We next trained a model from random initialized weights, which we call training \textit{from scratch}, using the training data from NSC1. This model (NSC1) resulted in better WER on NSC1 but very high WER on LS (46.82\% on dev-clean).

\begin{table}[th]
\centering
\caption{NSC1 - training from scratch vs fine-tuning, QuartzNet 15x5, greedy WER (\%)}
\label{tab:NSC1}
\scalebox{0.9}{
\begin{tabular}{c c c c c}
 \toprule
 {\textbf{Train}} &
 {\textbf{Fine-tune}} &
 \multicolumn{2}{c}{\textbf{LS}} & \multirow{2}{*}{\textbf{NSC1}} \\
 \textbf{dataset}& \textbf{dataset} & \textbf{dev-clean} & \textbf{dev-other} & \\
 \midrule
 LS    & - & 3.87 & 11.05 & 32.73 \\
 LS+CV & - & 3.99 & 10.89 & 26.85 \\
 5D    & - & 3.92 & 10.59 & 22.51 \\
 NSC1  & - & 46.82 & 58.83 & 18.07 \\
  \midrule
 5D   & NSC1 & 19.87 & 31.75 & 10.02 \\
 5D   & NSC1+LS & 4.24 & 11.47 & 9.53 \\
 \bottomrule
\end{tabular}
}
\end{table}

The next experiment was to pre-train on 5D and fine-tune on NSC1. This model shows a large reduction in WER reaching 10.02\% compared to a model trained on NSC1 only (WER 18.07\%). However, WERs on LibriSpeech are worse. When training only on the target dataset, TL leads to catastrophic forgetting on the datasets used for pre-training.

Finally, we observed the best WER on the target dataset with slightly worse WERs on pre-training datasets if we do transfer learning simultaneously on both target and pre-training datasets (see the last row of Table.~\ref{tab:NSC1}).

NSC2 is the more challenging dataset, because it has numerous non-English words (Singaporean local named entities). Table.~\ref{tab:NSC2} shows that fine-tuning on both NSC1 and NSC2 yields low WERs on both datasets simultaneously which is impossible if transfer learning is done on the target dataset only.

\begin{table}[th]
\centering
\caption{NSC2 - transfer learning, QuartzNet 15x5, greedy WER (\%)}
\label{tab:NSC2}
\scalebox{0.9}{
\begin{tabular}{c c c c}
 \toprule
 {\textbf{Train}} &
 {\textbf{Fine-tune}} &
  \multirow{2}{*}{\textbf{NSC1}} &  \multirow{2}{*}{\textbf{NSC2}} \\
 \textbf{dataset}& \textbf{dataset} &  & \\
 \midrule
 5D   & - & 22.51 & 54.10 \\
 \midrule
 5D   & NSC1 & 10.02 & 45.90 \\
 5D   & NSC2 & 72.89 & 6.02 \\
 5D   & NSC1+NSC2 & 10.61 & 6.67 \\
 \bottomrule
\end{tabular}
}
\end{table}

\subsection{African American Speech}

CORAAL \cite{coraal} is a public speech corpus of African American language. It consists of a set of interviews with African American speakers from Washington, D.C., rural North Carolina, and upstate New York. We created our own train/validation split by selecting 11 speakers for evaluation and setting aside the remaining samples for training.
The validation speakers were selected to be reasonably diverse, with both male and female speakers, speakers from all age and socioeconomic groups, and speakers from each location.
The resulting split has 88.6 hours of training utterances and 4.9 hours of validation utterances.

As this is a relatively small dataset, we first performed experiments using QuartzNet 5x3 - a small model with 6.4 million parameters. We  pre-trained two QuartzNet 5x3 models: one on the WSJ dataset only \cite{wsj} and another on CORAAL training samples only. In both cases we used 400 epochs for training. Then we fine-tuned the WSJ model (1) on CORAAL for an additional 100 epochs and (2) on a combination of CORAAL and WSJ, also for 100 additional epochs.

Next we experimented with the much larger QuartzNet 15x5 model. The model was pre-trained on the five datasets (5D) from Section 4.1 for 100 epochs, (2) trained on CORAAL only for 100 epochs (with a learning rate of 0.001 instead of 0.01)
, and (3) trained on 5D and fine-tuned on CORAAL or on CORAAL+WSJ for 100 additional epochs.
All models were evaluated on the WSJ dev-93 and eval-93 sets, and on our CORAAL validation set. Results are shown in Table~\ref{tab:Coraal}.

\begin{table}[th]
\centering
\caption{CORAAL - fine-tuned QuartzNet 5x3 and QuartzNet 15x5 have much better accuracy than models trained from scratch, greedy WER (\%)}
\label{tab:Coraal}
\scalebox{0.8}{
\begin{tabular}{l c c c c c}
 \toprule
  \multirow{2}{*}{\textbf{Model }} &
 {\textbf{Train}} &
 {\textbf{Fine-tune}} &
 \multicolumn{2}{c}{\textbf{WSJ}} & \multirow{2}{*}{\textbf{CORAAL}} \\
 &\textbf{dataset}& \textbf{dataset} &
 \textbf{dev-93} & \textbf{eval-92} & \\
 \midrule
  \textbf{5$\times$3} & CORAAL &- & 65.39 & 60.32 & 45.59\\
  & WSJ    & - &15.18 & 11.68 & 84.43 \\
  & WSJ    &  CORAAL & 45.39 & 38.12 & 41.48 \\
  & WSJ    & CORAAL+WSJ & 15.04 & 11.80 & 43.38\\
  \midrule
 \textbf{15$\times$5} & CORAAL & - & 80.98 & 77.30 & 60.15 \\
 & 5D & - &5.55 & 3.38 & 50.91 \\
 & 5D & CORAAL & 25.76 & 20.84 & 25.09 \\
 & 5D & CORAAL+WSJ & 8.34 & 5.57 & 25.92 \\
 \bottomrule
\end{tabular}
}
\end{table}

Once again, we saw that fine-tuning only on the target dataset (i.e. CORAAL) leads to catastrophic forgetting on the original datasets. In both the 15x5 and 5x3 models, training on CORAAL only results in a large increase in WER on the WSJ; however, with TL on both datasets, we can retain most of the accuracy on WSJ while suffering only a slight degradation to the CORAAL WER.
Additionally, we see that even though the dataset is very small, fine-tuning can still be done effectively with a larger model, even when only using CORAAL data.

\section{Cross-language transfer learning}

The common scheme of all experiments in this section is that we trained or fine-tuned with very limited amounts of  data compared to the 3,300 hours on which the English QuartzNet  model was trained. For German, Spanish and Russian  we perform two kinds of experiments: (1) training QuartzNet 15x5 and QuartzNet 5x3 from scratch, and (2) fine-tuning the pre-trained QuartzNet models on our target language.

\subsection{Datasets}

We used only the data from German, Spanish and Russian parts of Mozilla's Common Voice project \cite{ardila2019common}.
The datasets have been pre-processed by (1) converting from mp3 to the mono wav format sampled at 16 kHz and (2) processing transcripts to lower-case, removing punctuation and out-of-vocabulary characters.
We used the `train' and `dev' splits for training and validation correspondingly. Training sets for German, Spanish, and Russian consist of 118.5, 96, and 16 hours respectively. All WERs were measured on dev split of the data.

\subsection{Training and fine-tuning parameters}

The Russian model was trained for 512 epochs while the Spanish and German ones for 256 epochs. We used a batch size of 32 per GPU and trained on a single DGX with 8 V100 GPUs (therefore, total batch size is 32x8=256). We varied initial learning rate from 0.01 to 0.02 and from 0.001 to 0.002 with a cosine annealing learning rate policy and a warmup ratio of 12\% for all experiments. We found that the best learning rate for the fine-tuned model is about 10x smaller than the best one for the training from scratch model. NovoGrad \cite{novograd2019} optimizer was used in all experiments with weight decay in the range 0.001-0.002, $\beta_1=0.95$, and $\beta_2=0.25$.

\subsection{Results}

\begin{figure}[ht]
 \includegraphics[width=\linewidth]{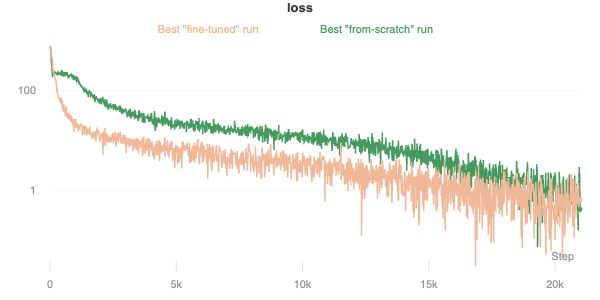}
 \caption{QuartzNet 15x5 training loss for Russian experiments.}
 \label{fig:mulitlang}
\end{figure}

Table \ref{tab:multilang} reports validation WER from the best runs only. The largest improvement between trained from scratch and fine-tuned experiments was achieved for the Russian language. This is likely because the Russian dataset is several times smaller than the German or Spanish datasets and therefore it should benefit the most from the fine-tuning schema. Note that even though the Russian from-scratch models failed to generalize, they easily fit the training data. Figure.~\ref{fig:mulitlang} demonstrates training loss plots for the best Russian fine-tuned and trained from scratch models. At the end, both models were able to fit the data acceptably. Therefore, starting from a pre-trained English encoder acted as a strong regularizer and improved convergence speed.

Since datasets for these languages are relatively small, we also experimented with the smaller QuartzNet 5x3. This model was pre-trained on WSJ  only. Overall, across all experiments, using this smaller model performed worse than fine-tuning a larger one (15x5) pre-trained on more data (see Table \ref{tab:multilang}).

\begin{table}[th]
\centering
\caption{ASR transfer learning for German (CV-ge), Spanish (CV-sp) and Russian(CV-ru) parts of Common Voice dataset. Training from scratch vs fine-tuning, greedy WER (\%)}
\label{tab:multilang}
\scalebox{0.8}{
\begin{tabular}{c c c c c}
\hline
 \toprule
 \multirow{2}{*}{\textbf{Language}} &
 \multirow{2}{*}{\textbf{Model}} &
 \textbf{Train from} &
 \textbf{Fine-} &
 \multirow{2}{*}{\textbf{WER,\%}} \\
 &&\textbf{scratch} & \textbf{tune}& \\
 \midrule
\multirow{4}{*}{German}  & \multirow{2}{*}{ 5x3}  & CV-ge                           & -                   & 30.42              \\
                         &                          & WSJ                           & CV-ge             & 28.61              \\ \cline{2-5}
                         & \multirow{2}{*}{ 15x5} & CV-ge                            & -                   & 23.35          \\
                         &                          & 5D                            & CV-ge              & \textbf{18.65} \\ \hline
\multirow{4}{*}{Spanish} & \multirow{2}{*}{ 5x3}  & CV-sp                           & -                   & 23.29          \\
                         &                          & WSJ                           & CV-sp             & 22.93          \\ \cline{2-5}
                         & \multirow{2}{*}{ 15x5} & CV-sp                            & -                   & 19.82          \\
                         &                          & 5D                            & CV-sp             & \textbf{14.96} \\ \hline
\multirow{4}{*}{Russian} & \multirow{2}{*}{ 5x3}  & CV-ru                           & -                   & 42.18          \\
                         &                          & WSJ                           & CV-ru             & 40.24          \\ \cline{2-5}
                         & \multirow{2}{*}{ 15x5} & CV-ru                            & -                   & 59.50          \\
                         &                          & 5D                            & CV-ru             & \textbf{32.20} \\ \hline
\end{tabular}
}
\end{table}

\section{Cross domain ASR adaptation}

In this section, we investigate transfer learning for domain adaptation using a large dataset from the financial domain.

\subsection{Dataset}

The proprietary financial dataset was compiled by Kensho and comprises over 50,000 hours of corporate earnings calls, which were collected and manually transcribed by S\&P Global over the past decade. The calls are conducted in English and contain a large variety of accents and audio quality. The style of speech consists of narrated presentations as well as spontaneous question/answer sections.
In contrast to the 5D dataset, the vocabulary of the Kensho corpus contains a wide variety of specialized terms including domain-specific terminology, acronyms and product names. This shift in word distribution, as well as the large volume of data, make for an ideal experiment in large-scale domain adaptation.

\subsection{Results}

Experiments were performed using 512 GPUs, with a batch size of 64 per GPU, resulting in a global batch size of 512x64=32K. All  other hyper-parameters were kept consistent with experiments in earlier sections. Similar to the previous sections, we compare a model trained from scratch versus a fine-tuned model that was pre-trained on the 5D dataset.

\begin{figure}[htb!]
 \centering
 \includegraphics[width=0.9\linewidth]{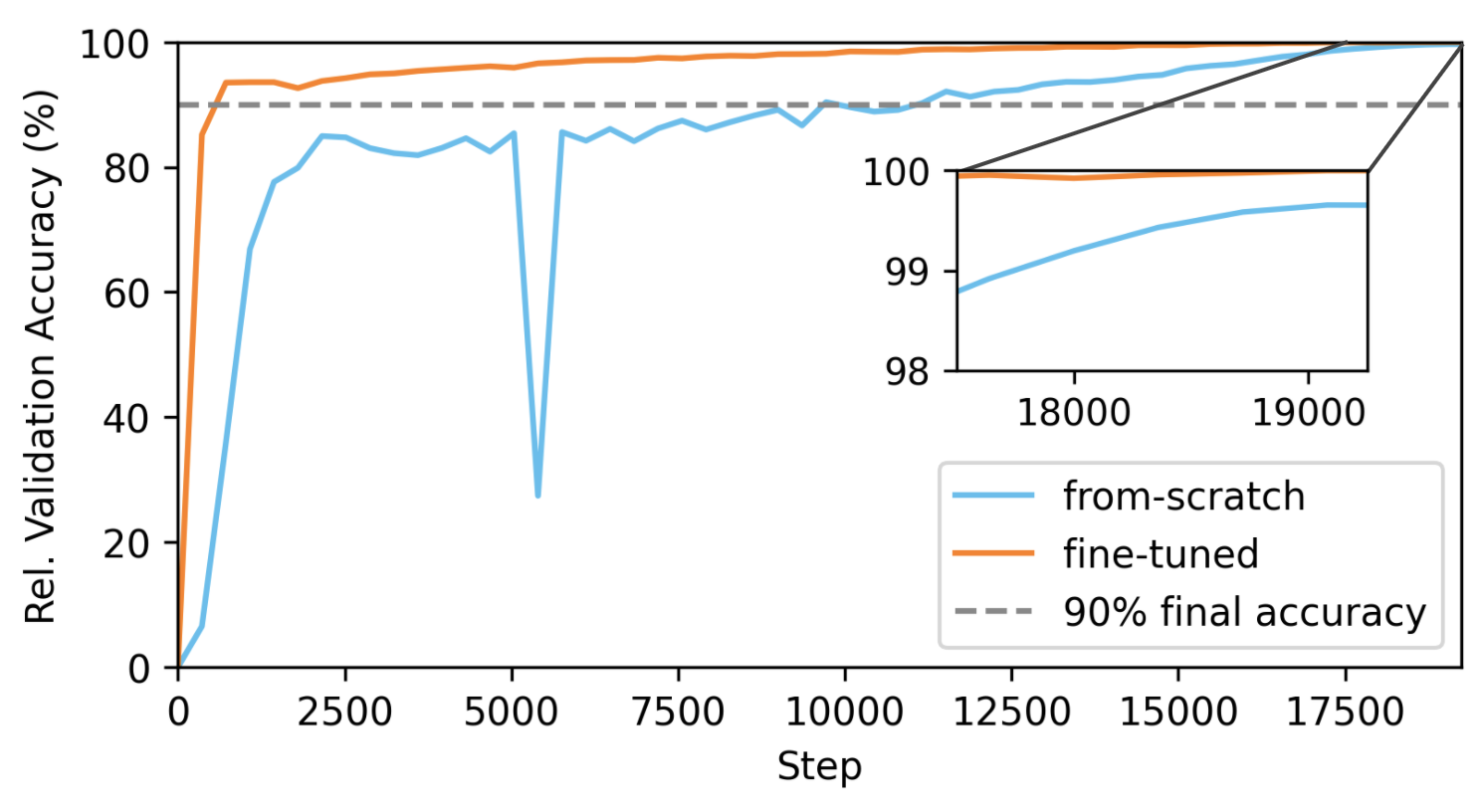}
 \caption{Relative validation accuracy during training on financial data using a model trained from scratch (blue line) and a pre-trained model (orange line). Dashed line indicates 90\% of final accuracy level as a guide to the eye.}
 \label{fig:domain_ft}
\end{figure}
The results in Figure.~\ref{fig:domain_ft} demonstrate consistently higher accuracy throughout the training process when using a pre-trained model, as well as a better optimum when fully converged. Accuracy is normalized to the best measured value, achieved by the pre-trained model at the end of the full training procedure.

In the large data limit, it may be favorable to reduce training time at the cost of final accuracy. To quantify this, we observe that the pre-trained model attains 90\% of final relative accuracy approximately 18x faster than a model trained from scratch.  (see Figure.~\ref{fig:domain_ft}, dashed line)

\section{Conclusions}

A transfer learning approach based on reusing a pre-trained QuartzNet network encoder turns out to be very effective for various ASR tasks. This general approach worked across multiple target domains: (1) transfer learning within English language to different accents, (2) cross-language transfer learning, and (3) cross domain adaptation. In all our experiments, we observed that fine-tuning a good baseline always performed better than training from scratch. Our approach worked well both in experiments where training data was very small and very large.
It is preferred to fine-tune large, more accurate models even if the fine-tuning dataset is small. Starting from a good model acted as a strong regularizer and dramatically improved generalization. In a very large data setting it significantly sped up the convergence while still producing the best final performance.

\section{Acknowledgments}
We would like to thank NVIDIA AI Applications team for the help and valuable feedback as well as Keenan Freyberg for facilitating collaboration between NVIDIA and Kensho.

\bibliography{bibliography}

\end{document}